\documentclass[aps, nofootinbib, twocolumn,groupedaddress,floatfix,nofootnotesinbib,preprintnumbers]{revtex4}
\usepackage{graphicx}
\usepackage{epsfig}
\usepackage{epstopdf}
\usepackage{rotating}
\usepackage[mathscr]{eucal}
\usepackage{amsmath,amssymb,bm,graphics}
\usepackage[letterpaper=true]{hyperref}

\let\mathbf=\bm

\newif\ifarxiv
\arxivfalse

\begin{document}
\def\Nfour{\mathcal N\,{=}\,4}
\def\Ntwo{\mathcal N\,{=}\,2}
\def\Nc{N_{\rm c}}
\def\Nf{N_{\rm f}}
\def\x{\mathbf x}
\def\q{\mathbf q}
\def\f{\mathbf f}
\def\v{\mathbf v}
\def\C{\mathcal C}
\def\w{\omega}
\def\vs{v_{\rm s}}
\def\S{\mathcal S}
\def\half{{\textstyle \frac 12}}
\def\twothirds{{\textstyle \frac 23}}
\def\third{{\textstyle \frac 13}}
\def\t{\mathbf{t}}
\def\T{\mathcal {T}}
\def\O{\mathcal{O}}
\def\E{\mathcal{E}}
\def\p{\mathcal{P}}
\def\H{\mathcal{H}}
\def\uh{u_h}
\def\R{\ell}
\def\Ro{\chi}
\def\del{\nabla}
\def\eps{\hat \epsilon}
\def\nn{\nonumber}
\def\K{\mathcal K}
\def\inf{\epsilon}
\def\cs{c_{\rm s}}
\def\A{\mathcal{A}}
\def\e{{e}}
\def\r{{\xi}}
\def\x{{\mathbf x}}
\def\w{{w}}
\def\rr{{\xi}}
\def\uo{{u_*}}
\def\u{{\mathcal U}}
\def\G{\mathcal{G}}
\def\Deltax{\Delta x_{\rm max}}
\def\L{{\bm L}}

\title
{Extra dimensions, black holes and fireballs at the LHC}

\author{Anastasios Taliotis}

\affiliation
   {Theoretische Natuurkunde, Vrije Universiteit Brussel
and The International Solvay Institutes Pleinlaan 2, B-1050 Brussels, Belgium}

\date{December 22, 2012}

\begin{abstract}

\hspace{-0.17in} The collision of two gravitationally interacting, ultra-relativistic, extended sources is being examined. This investigation classifies the transverse distributions that are collided for fixed collision energy, according to whether one or two (a small and a large) apparent horizons may or may not be formed in a flat background in 4 dimensions. The study extends to the thermodynamical properties of the objects that are created, which exhibit a universal behavior in their entropy, and, suggests the elimination of the possibility in observing black holes (BHs) at the LHC in the absence of extra dimensions. On the other hand, including extra dimensions, and assuming that the matter is localized (dense) enough in those directions, opens new avenues in creating BHs at energies of the order of TeV. The investigation is carried further to $AdS_5$ backgrounds and makes connections with the implications for the quark-gluon plasma (QGP) formation in heavy ion collisions. In particular, classes of the geometries found suggest that a BH is formed if and only if the (central collision) energy is sufficiently large compared to the transverse scale of the corresponding gauge theory side stress-tensor. This implies that  when the scattering in the gravity description is mapped onto a heavy ion collision problem yields a result, which is in accordance with the current intuition and data: QGP is formed only at high enough energies compared to $\Lambda_{QCD}$, even for central processes. Incorporating weak coupling physics and in particular the Color Glass Condensate (CGC) model, a satisfactory fitting with the RHIC and the LHC data for multiplicities may be established.

\end{abstract}


\pacs{}

\maketitle
\iftrue
\def\thefootnote{\fnsymbol{footnote}}
\footnotetext[1]{Email: \tt atalioti@vub.ac.be}
\def\thefootnote{\arabic{footnote}}
\fi

\parskip	2pt plus 1pt minus 1pt

\section{Introduction}

\vspace{-0.12in}

Gravitational ultra-relativistic collisions of two shock-waves generated by point-like sources on the transverse plane have been discussed in several contexts, \cite{Aichelburg:1970dh, D'Eath:1976ri,D'Eath:1992hb,D'Eath:1992hd,D'Eath:1992qu,Coelho:2012sy,Coelho:2012sya,Herdeiro:2011ck,Amati:1993tb,Kohlprath:2002yh,Amati:2007ak,Veneziano:2008xa,Ciafaloni:2008dg,Eardley:2002re,Sfetsos:1994xa,Constantinou:2011ju,Gal'tsov:2009zi,Mureika:2011hg}.  Earliest studies \cite{Eardley:2002re} have shown that such collisions typically create a single black hole (BH) given sufficiently large energy compared to the impact parameter involved. In particular, for central collisions, these and some follow up works \cite{Eardley:2002re,Kovchegov:2009du,Lin:2009pn,Gubser:2008pc,Arefeva:2012ar,Kiritsis:2011yn}, have shown that a BH is formed for any (small) collision energy. Works \cite{Kohlprath:2002yh,Casadio:2012ng} generalized the analysis to extended distributions 
and found the conditions under which a single BH may be formed. 

In this work further steps are taken in that direction. We classify the energy-momentum (transverse) distributions that are collided according to (i) always a trapped  horizon\footnote{Unless otherwise specified, we will use the terminology trapped horizon, trapped surface and BH interchangeably.} is formed independently on the amount of the collision energy (as in \cite{Eardley:2002re}), (ii) one trapped horizon is formed for sufficiently large energy (as in \cite{Kohlprath:2002yh}) (iii) two (and sometimes more) trapped horizons, one small and one large, are formed for sufficiently large energy. The classification is done in the two cases of flat backgrounds and AdS$_5$ backgrounds where the creation of mini BHs (see also \cite{Kiritsis:2011qv}) in the accelerators and the implications in heavy ion collisions are discussed respectively. We restrict the classification to transversally symmetric geometries \footnote{The transverse symmetry will be defined rigorously for each background. Other than this symmetry and three mild and physical assumptions, the classification is completely generic.} in the spirit of trapped surfaces. Our analysis assumes a zero impact parameter limit as this case suffices in order to illustrate our claims.

{\it BHs from shock-wave collisions is not a new concept. What is new here is:}

\vspace{0.1in}

1. Category (iii) of trapped surfaces formed under a dynamical process has not been studied so far in the literature and is one of the main results of our work. In fact, evidence of creation of small (in addition to large) BHs in flat backgrounds has not been reported up to today. The underlying set-ups consist of examples of the (known) non-uniqueness of the solutions regarding the trapped surface equations (see \cite{Eardley:2002re}). 

\vspace{0.07in}

2. The small BHs seem to have a curious property: their size (and hence their entropy) decreases as the energy increases and this behavior is exactly opposite to that of the large BHs. If this was an equilibrium state\footnote{In fact it is not; this is the beginning of a very violent process.} it would imply a negative temperature (T).  There are a number of interesting scenarios to be discussed.  An attractive explanation is that the resulting trapped surface, having two (apparent) horizons, will evolve into a Reissner-Nordstrom-like (RN) of BH whose inner and outer horizon share a similar behavior, as the energy varies, with the corresponding small and large apparent horizons that we find in our analysis. In fact, there exists a critical energy such that the two surfaces have the same size, which by analogy to the RN BH, it could be the extremal limit. Lowering the energy further, in our case we find that no trapped horizons may be formed. In the RN case, it yields a naked singularity, which violates the ``cosmic censorship" conjecture. The role of the corresponding scale of the electric field in the RN solution, is (possibly) played by the transverse width of the colliding distributions (see also \cite{Nicolini:2008aj,Nicolini:2011dp}). In particular, working in an AdS$_5$ background, such a configuration could model \cite{Kovchegov:2009du,Lin:2009pn,Gubser:2008pc,Arefeva:2012ar,Kiritsis:2011yn,Albacete:2008vs,Albacete:2009ji,Taliotis:2010pi} heavy ion collisions. As the inner horizon has a different thermodynamical behavior than the outer region, an RN-like scenario could possibly describe a hotter fireball in the center of the produced plasma, which absorbs heat from the surrounding shell, and a thermodynamically unstable periphery \cite{Gubser:2000ec,Konoplya:2008rq,Chamblin:1999hg,Louko:1996dw} that undergoes a hadronic phase transition.

\vspace{0.07in}

3. In the high energy (E) limit, the entropy (S) of the (large) BHs for all the three classes of the colliding distributions and for both, flat and AdS backgrounds, exhibits a universal behavior. The entropy as a function of the c.m. energy $s=4E^2$ and the transverse scale $k$ of the distribution behaves as $S\sim k^2s $ and $S\sim (s/k^2)^{1/3}$ for flat and AdS$_5$ backgrounds respectively independently on the details of the profiles of the (bulk) distributions that are collided. In general, k can be $s$ dependent (e.g. see (\ref{Ns})). In fact, this universal behavior is independent on the impact parameter and any extra dimensions (if these ingredients are introduced) as long as the energy is sufficiently large. Applying the AdS/CFT results of this work in heavy ion collisions and incorporating weak coupling physics, a satisfactory fitting with the total multiplicities data may be achieved. This can be seen in fig. \ref{N}.

\vspace{0.07in}

4. Under the assumption that extra dimensions do not exist, this work motivates that the LHC is a ``black hole safe" machine. The parameter that controls the BH formation is the dimensionless combination  $Ek/M_p^2 \geq w$ where $E$ is the collision energy, $k$ the is transverse scale associated with the colliding distribution $\rho$ and can be taken of the order of $\Lambda_{QCD}$, $M_p$ is the Planck mass and $w$ is a dimensionless number that is predicted from the trapped surface equations and naturally depends on $\rho$. Usually $w$ is of the order of 1 or less. In particular, for the boosted Woods-Saxon profile for nuclear distributions, which is a suitable phenomenological model for collisions at LHC, $w\approx 0.4$. This implies that the collision energy that is required to create BHs at the LHC is way beyond the machine's possibilities.
Including extra dimensions on the other hand, could change things dramatically. Towards the end of the conclusions, it is argued that BHs could be created even at TeV scales provided that the distributions that are collided are sufficiently dense in the compact directions. In other words, the underlying assumption is that the matter is confined almost completely on the 3-brane.

\vspace{0.07in}

5.  Classes (ii) and (iii) for AdS backgrounds, advocate the idea that not any distribution of matter at low energy when collided, yields a BH. Mapping this idea, through the AdS/CFT \cite{Maldacena:1997re} dictionary  \cite{Witten:1998qj,Gubser:1998bc,deHaro:2000xn}, to heavy ion collisions as in \cite{Kang:2004jd,Giddings:2002cd,Gubser:2008pc,Gubser:2009sx,Lin:2009pn,Kiritsis:2011yn,Kovchegov:2007pq,Kovchegov:2009du,Aref'eva:2009wz,Arefeva:2012ar,AlvarezGaume:2008fx} implies that QGP \cite{Adams:2003im} is formed provided that the $E$ of the colliding (dual picture) glue-balls is sufficient. In particular, according to our calculations, $E$ must be much larger than the transverse scale of the corresponding gauge-theory stress tensor (that can be taken of the order of $\Lambda_{QCD}$), even for a zero impact parameter collision. This is a totally new result and extends the results of the existing literature, which state that QGP, under a holographic investigation, is always formed for any small collision energy as long as the collision is (perfectly) central . In particular, as we show for AdS backgrounds, the parameter controlling the BH formation appears in the combination $E/k \geq w$ (contrast with point 4. above for flat backgrounds) where $w$ depends on the particular (bulk, 5-dimensional) stress-tensor distribution. Central nuclei collisions at energies of a few eV's should not result to a QGP formation;  this has been the initial motivation of this part of our investigation.

\vspace{0.1in}
We organize this paper as follows.

In section \ref{method} we set up the problem and we briefly review the method that we will use in order to support our claims.

Section \ref{3} deals with the classification of the energy momentum distributions for flat backgrounds. Three classes of distributions are found and one example for each one is given. In addition, the thermodynamical consequences, the final product of the collision, as well as 
the implications to collisions at the LHC are discussed.

Next section, section \ref{4}, generalizes the discussion of section \ref{3} to AdS backgrounds and gives particular emphasis to the consequences in the applications of the gauge/gravity duality in heavy ion collisions.

Finally, in section \ref{conc} we summarize our findings and suggest possible extensions of this analysis. In particular, we include an extended discussion for extra flat dimensions and argue that the inclusion of such compact directions could make BHs creation at the LHC feasible. 


\section{Setting up the problem}\label{method}
%
In this section we review very briefly the trapped surface method that is required for our subsequent analysis. More details can be found in \cite{Eardley:2002re,Kovchegov:2009du,Lin:2009pn,DuenasVidal:2012sa,Gubser:2008pc,Arefeva:2012ar,Kiritsis:2011yn} while extended notes exist in the appendices of \cite{Kiritsis:2011yn}.

We begin with the shock metric in a flat background created by a fast-moving source moving along the $-x^3$ direction
\vspace{-0.1in}
\begin{align}\label{ds1}
&ds^2 = -2 dx^+ \, dx^- + \phi
(x^1,x^2)\delta(x^+)  d x^{+ \, 2} + d {\bf x_\perp}^2\notag\\&
 {\bf x_{\perp}}=(x^1,x^2) 
\end{align}
where $\phi$ is the transverse profile of the shock and has dimensions of length, $x^{\pm}$ are light-cone coordinates (defined as $x^{\pm}=(x^0\pm x^3)/\sqrt{2}$) and $\delta$ is the Dirac's function. This ansatz satisfies trivially all the Einstein's equations except the $(++)$ component, which yields
\vspace{-0.1in}
\begin{align}\label{Ein}
R_{++} & = 8\pi G \left( T_{+ +} - \frac{1}{2} \, g_{++} T_{\mu}^{\mu}\right)
 =>-\delta(x^+)\frac{\nabla_{\perp}^2}{2} \phi \notag\\&
 = 8\pi GT_{++}  
 =EG\rho \delta(x^+), \hspace{0.15in}T_{++}\equiv E\rho \delta(x^+)
\end{align}
allowing one to identify the stress energy tensor component $T_{++}$ with the transverse Laplacian of $\phi$ (times $\delta(x^+)/G$)\footnote{From now on we drop the dimensionless proportionality factors for simplicity as they do not alter qualitatively the final results.}. In equation (\ref{Ein}) we find it convenient to define $T_{++}$ by pulling out its total energy $E$ and its delta function dependence; this defines the density $\rho=\rho({\bf x_{\perp}})$.

The trapped surface consists of two pieces, $S_{+}$ and $S_-$. These are parametrized with the help of two functions, $\psi_+$ and $\psi_-$, which satisfy the following differential equation
\begin{align}\label{de}
 \nabla^2_{\perp} (\psi_{\pm}-\phi_{\pm})=0.
\end{align}
Obviously, $\nabla^2_{\perp}  \phi_{\pm}$ provides a source term for  $\nabla^2_{\perp}  \psi_{\pm}$. The missing ingredient is the boundary conditions. Dropping the indices $\pm$ from $\psi_{\pm}$ from now on assuming a zero impact parameter and identical shocks we have $\psi_{+}=\psi_{-}=\psi$. The boundary conditions then read
 \begin{align}\label{BC}
\psi \Big |_C=0 \hspace{0.4in}  \sum_{i=1,2}\left[ \nabla_{x^i} \psi  \nabla_{x^i} \psi  \right]  \Big |_{C}=8
 \end{align}
 for some curve $C$, which defines the boundary of the trapped surface (so the curve $C$ is the trapped horizon) and where both, $S_+=S$ and $S_-=S$ end. The set of equations, equations (\ref{de}) and (\ref{BC}) that define the trapped surface are generalized for AdS spaces in a straightforward manner. In particular, the differential equation (\ref{de}) for AdS$_5$ is replaced by 
 \begin{align}\label{ads}
 (\Box_{\perp}-3/L^2)(\psi_{\pm}-\phi_{\pm})
 \end{align}
where $\Box_{\perp}$ is the scalar operator in $AdS_3$ while the (left hand side of the) second boundary condition in (\ref{BC}) should be contracted with the metric tensor of AdS$_3$.

\section{Flat Background BHs production}\label{3}
\subsection{General classification}\label{3A}

The solution to the system of equations (\ref{de}) and (\ref{BC}) is 
\begin{align}\label{gc1}
\psi(x_{\perp})=\phi(x_{\perp})-\phi(x^c_{\perp})
\end{align}  
where the curve $C: x_{\perp}=x_{\perp}^c$ defines the trapped horizon, which is a 2-dimensional disk. The second boundary condition then specifies $x^c_{\perp}$ in terms of the data of the shock yielding\footnote{All the positive constants will be set equal to one from now on.}

\begin{align}\label{e7}
(\psi(x^c_{\perp})')^2=8=> \phi(x^c_{\perp})' \sim 1.
\end{align}  
Now taking into account that equation $\nabla_{\perp}^2\phi_{x_{\perp}} \sim EG \rho$ (see equation  (\ref{Ein})) implies $1/x_{\perp}\left(x_{\perp}\phi'\right)'\sim EG \rho$, then equation (\ref{e7}) yields
\begin{align}\label{con1}
\frac{\int_0^{x_{\perp}^c}x\rho(x,k) dx}{x_{\perp}^c} \sim \frac{1}{GE}=>\frac{\int_0^{x_{\perp}^c}x\rho(x,k) dx}{kx_{\perp}^c} \sim \frac{1}{GEk}
\end{align}  
where $k$ a transverse scale whose inverse determines the width of the distribution $\rho$. Equation (\ref{con1}) is the basic equation of the flat background investigation and its consequence  can be fit in three classes depending on $\rho$. The only assumptions for $\rho$ is that it must be (i) everywhere positive (in space-time) in order to be a physical energy distribution\footnote{This is the weak (and since $T_{\mu}^{\mu}=0$ it is also the strong) energy condition: $T^{\mu \nu}n_{\mu}n_{\nu}=T^{--}(n_-)^2>0$ for any time-like (and space-like) vector $n_{\mu}$ if and only if $T^{++}=T_{--} \sim \rho>0$.}, (ii) integrable in order to yield a finite total collision energy and (iii) it must have a monotonic fall-off in the sense $d/dx(x\rho(x))=0$ has only a single real root\footnote{By assumption as one expects that physical distributions either will increase, reach a maximum value and then decrease at infinity or they will start from a maximum value (which can be infinite but integrable at the origin) and keep decreasing. Multiplying by the measure x (of polar coordinates as the system is effectively 2-dimensional), turns either type of the distribution to a bump function, which vanishes at zero and at infinity and it has a maximum between. If this assumption is removed, then as we will see, more than two trapped horizons are allowed as the number of trapped surfaces increases with the number of the extrema of the left hand side of (\ref{con1}). In particular, assumption (iii) restricts the number of extrema to one or less (this is not a straightforward fact-see footnote in subsection \ref{3A3}).}.

In particular, assumption (ii) implies that the numerator in the left hand side of (\ref{con1}) is finite as $x_{\perp}^c \rightarrow \infty$. Assumptions (i) and (ii) combined, imply that the roots of the left hand side of (\ref{con1}) is at $x_{\perp}^c=\infty$ and possibly at $x_{\perp}^c=0$ depending on the behavior of $\rho$ at the origin. As it will be seen shortly, the behavior of $\rho(x_{\perp})$ near $x_{\perp}=0$ partitions the set of positive and integrable $\rho$'s into three disconnected classes according to the details of the trapped horizons. 

The physical meaning of formula (\ref{con1}) is that given a total collision energy $E$, a trapped horizon exists if and only if there exists an $x_{\perp}^c$ such that the energy that is enclosed inside the circle of radius $x_{\perp}^c$ is equal (modulus constant pre-factors) to $x_{\perp}^c/G$.
An observation for (\ref{con1}) comes about when notting that the left hand side of the equation (after the $=>$ symbol) is a dimensionless quantity of the variable $kx_c$. This implies that the dimensionless quantity that determines the existence and the behavior of the trapped surface(s) in a 4-dimensional flat backgrounds is set by the product of two energy scales as $E\times k$ in units of $G$. This fact should be contrasted with the next section, which deals with AdS$_5$ backgrounds. There, as will see, the behavior of the corresponding dimensionless coupling is dramatically different and interesting. 

\vspace{-0.15in}

\subsubsection{Always a single black hole for any energy }\label{3A1}

This case corresponds to the distributions, which near the origin behave as $\rho(x_{\perp})\sim 1/x_{\perp}^n+$(sub-leading) where $2>n>1$ \footnote{ $\rho$ can not drop faster than $1/x^2$ because  the total energy of the collision is finite by assumption.} \footnote{In particular, this implies that $(x\rho)'$ can not become zero in the interval $(0,\infty)$ because $x\rho$ attains its unique (due to assumption (iii)-see below equation (\ref{con1})) extremum at $x=0$.}. This implies that the left hand side of (\ref{con1}) diverges at $x_{\perp}^c=0$ and goes to zero as  $x_{\perp}^c \rightarrow \infty$ and as a result, there is always a single\footnote{Uniqueness can be shown using the fact of the previous footnote and working analogously to the corresponding proof of section \ref{3A3} (see the proof in the relevant footnote of that section).} solution for any $E$ to equation (\ref{con1}) for some $x^c_{\perp}$.
\begin{figure}
\includegraphics[scale=0.62]{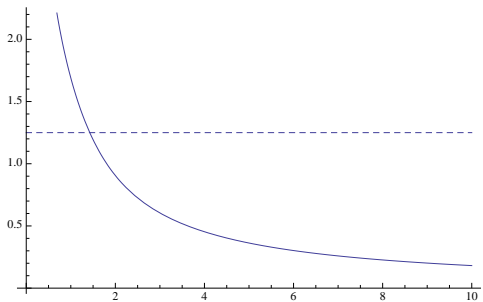}
\caption
    {\label{I1}
     The function $(\Gamma(1/4)-\Gamma(1/4,y^2)/y$ as a function of $y\equiv k x_{\perp}$. The dashed line corresponds to a given energy $E$. This is an example of a distribution that always results to a single trapped horizon regardless from the (small) value of $GkE$. }
\end{figure} 
As an example, we consider the (everywhere positive and integrable) distribution $\rho \sim k^2e^{-k^2x_{\perp}^2}/(kx_{\perp})^{3/2}$, which (for completeness) results to the shock $\phi \sim EG \sqrt{kx_{\perp}} \hspace{0.05in} _2F_2(\{1/4,1/4\};\{5/4,5/4\};-(kx_{\perp})^2)$ where $_pF_q$ the generalized hypergeometric function. Here $k$   is some transverse scale that fixes the width of the distribution. The  left hand side of (\ref{con1}) yields $\sim (\Gamma(1/4)-\Gamma(1/4,(k x_{\perp}^c)^2))/x^c_{\perp}$ ($\Gamma$'s are Gamma and incomplete Gamma functions). Last expression is plotted in fig. \ref{I1} as a function of $k x_{\perp}^c$ along with a constant line (see dashed line in the figure) of $1/(GkE)$. The intersection point $kx_{\perp}^c$ is the rescaled radius of the unique apparent horizon for the given values of $E$ and $k$.

It is evident that regardless from the value of $GkE$, which can be thought as a measure of the energy density on the transverse plane, a trapped horizon always exists: no mutter how dilute the energy is distributed on the transverse plane, a single BH will be formed.

\subsubsection{Marginal case: one black hole for sufficiently large energy}\label{3A2}
This case corresponds to distributions that  near the origin behave as $\rho(x_{\perp})\sim 1/x_{\perp}+$(sub-leading) \footnote{Since $(x\rho)'|_{x=0}=0$, it implies that $x\rho$ does not have any other extrema (due to assumption (iii)-see below equation (\ref{con1})).}. %
This implies that the left hand side of (\ref{con1}) goes to a constant as $x_{\perp}^c \rightarrow 0$ and goes to zero as  $x_{\perp}^c \rightarrow \infty$ and as a result, there is a single solution\footnote{Uniqueness can be shown using the fact of the previous footnote and working analogously to the corresponding proof of section \ref{3A3} (see the proof in the relevant footnote of that section).} $x^c_{\perp}$   to equation (\ref{con1}) but only for sufficiently large $E$.
\begin{figure}
\includegraphics[scale=0.62]{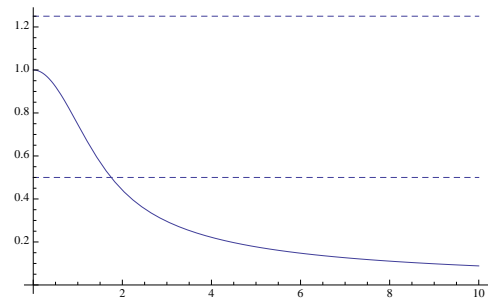}
\caption
    {\label{I2}
     The function Erf$(y)/y$ as a function of $y\equiv k x_{\perp}$. The dashed lines correspond to lines of constant energy $E$. This is an example of a distribution that results to a single trapped horizon but only for sufficiently small values of $1/GkE$ and hence sufficiently large values of $E$ in units of 1/kG. In particular, the top dashed line does not correspond to a trapped horizon formation while the bottom line does.}
\end{figure} 

As an example, we consider the (everywhere positive and integrable) distribution $\rho \sim k^2e^{-k^2x_{\perp}^2}/(kx)$, which (for completeness) results to the shock $\phi \sim GE kx_{\perp} \hspace{0.05in} _2F_2(\{1/2,1/2\};\{3/2,3/2\};-(kx_{\perp})^2)$. The  left hand side of (\ref{con1}) yields $\sim $Erf$(k x_{\perp}^c)^2))/x^c_{\perp}$ where Erf is the error function. Previous expression is plotted in fig. \ref{I2} as a function of $k x_{\perp}^c$ along with lines of constant values of $1/(GkE)$. The intersection point $kx_{\perp}^c$, if it exists, is the rescaled radius of the unique apparent horizon. It is evident that a single BH will be formed if and only if the transverse energy density is large enough.  

An interesting limit of this example is the limit where $k\rightarrow \infty$. Then, the distribution becomes a delta function. This is precisely the case of Eardley and Giddings \cite{Eardley:2002re} where they found that a BH is always formed. This limit is easily recovered by considering the ordering of limits  $\lim_{k \to \infty}[\lim_{x_{\perp} \to 0}$Erf$(k x_{\perp})/x_{\perp}]$, which is consistent with the ordering of limits that a delta function is defined and, which yields  $\sim \lim_{k \to \infty} k$. This implies that the intersection point of  the left hand side of (\ref{con1}) with the vertical axis (see fig. \ref{I2}) is shifted to infinity. Equivalently, $\lim_{k \to \infty} $Erf$(k x_{\perp}) /x_{\perp} \sim 1/x_{\perp}$ which is the \cite{Eardley:2002re} case and hence a BH is always formed in this limit. Effectively, this limiting case transforms into the previous class of geometries/distributions we have already studied.

\subsubsection{Two black holes for sufficiently large energy}\label{3A3}
Finally, the third class concerns distributions that behave as $\rho(x_{\perp}) \sim 1/x_{\perp}^n$+(sub-leading) where $n<1$. Then the left hand side of (\ref{con1}) vanishes at $x_{\perp}^c=0$ and since it also vanishes at $x_{\perp}^c=\infty$, it implies that given sufficiently large energy, there exist two roots for equation (\ref{con1}) resulting to two BHs: a small and a large one. The possibility of more than two roots of (\ref{con1}) and hence more than two BHs is excluded as a consequence of assumption (iii) in section \ref{3A}
\footnote{Pf: The extrema of the left hand side of (\ref{con1}) are the zeros of the function $f(x)=\rho x^2-\int_0^x \rho x dx$ and are located in the open interval $(0,\infty)$. We claim that $f$ begins from zero (because of assumption (ii)), then for small arguments it increases at positive values (because at small $x$, $\rho(x)$ increases slower than $1/x$) and it eventually asymptotes to a negative value (because of assumption (ii) which in particular implies that $\rho$ decays faster than $1/x^2$ at large $x$.). This implies that $f$ must have at least a (positive valued) maximum and at least one root. We claim this root is unique. To show this, it suffices to show that the maximum of $f$ is the unique extremum, which means that $f'$ must have a single root. Computing $f'$ we obtain $f'=x(x\rho)'$. But $f'\sim(x\rho)'$ has only one root in the open interval $(0,\infty)$ which is true by assumption.}.

As an example, we consider the (everywhere positive and integrable) distribution $\rho \sim k^2e^{-k^2x_{\perp}^2}$. The corresponding shock is $\phi \sim E G\left(Ei(k^2x_{\perp}^2)-\log(k^2x_{\perp}^2)\right)$ where the logarithm removes the logarithm of the exponential integral function $Ei$ (when expanded at small arguments) and hence it removes the delta function source keeping only the gaussian one.
\begin{figure}
\includegraphics[scale=0.62]{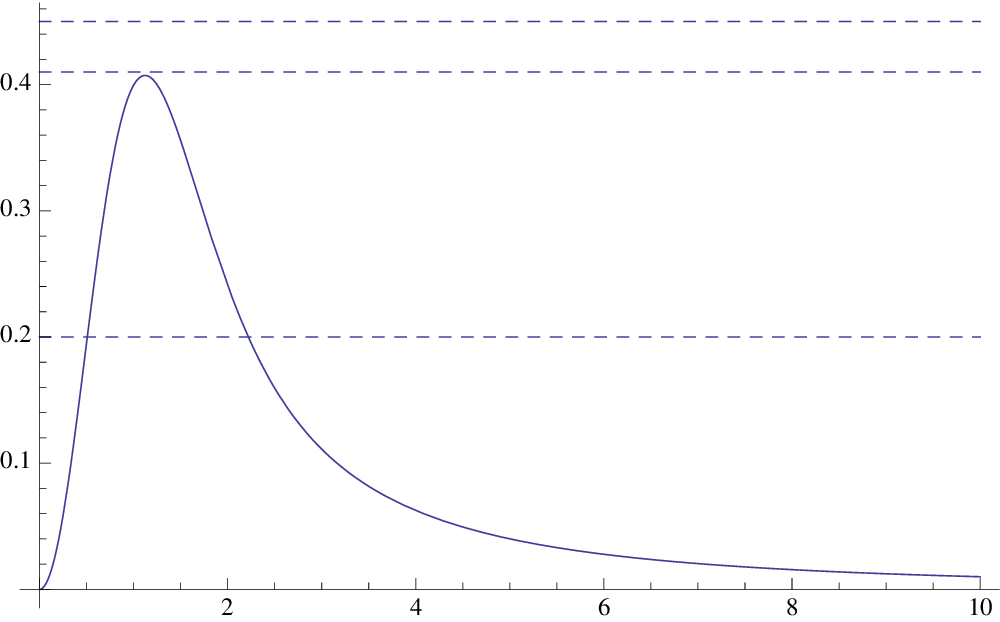}
\caption
    {\label{TSd}
     The function $(1-e^{-y^2})/y$ as a function of $y\equiv k x_{\perp}$. This is an example where there exist two trapped horizons provided that the transverse energy distribution is dense enough. The dashed lines correspond to lines of constant $E$. The top and bottom lines imply no BH and two BHs formation respectively. The middle line suggests that there is a critical energy such that the large and the small BH merge.}
\end{figure} 


The  left hand side of (\ref{con1}) yields $(1-e^{-k^2 x_{\perp}^2})/x_{\perp}$ and is plotted in fig. \ref{TSd} as a function of $k x_{\perp}^c$ along with lines of constant values of $1/(GkE)$. The two intersection points $kx_{\perp}^c$, if they exist, are the two (rescaled) radius of the apparent horizons. Evidently, there are two trapped surfaces, a small and a large one, if and only if there exist enough energy in the collision. In fact, the small trapped surface becomes smaller as $E$ increases with opposite behavior for the large one. What is more, is that there exists a critical value of the energy where only one trapped horizon exists while for even smaller energies, trapped horizons cease to exist at all. These features point, by analogy, to the direction of an extremal charged BH and a naked singularity situation respectively  (compare with \cite{Biswas:2011ar}). We will return on this in section \ref{3B}.

\subsubsection{Many-many apparent horizons}\label{3A4}
\begin{figure}
\includegraphics[scale=0.62]{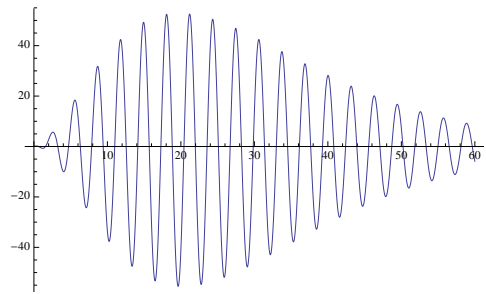}
\caption
    {\label{VC}
The function $x(x\rho)'$ as a function of $x$ (in units of $k$) where $\rho \sim k^2 \left( kx \cos^2(kx)\exp[-0.1kx]\right)$. This distribution clearly violates condition (iii) and more than two apparent horizons are thus expected.}
\end{figure}

\begin{figure}
\includegraphics[scale=0.62]{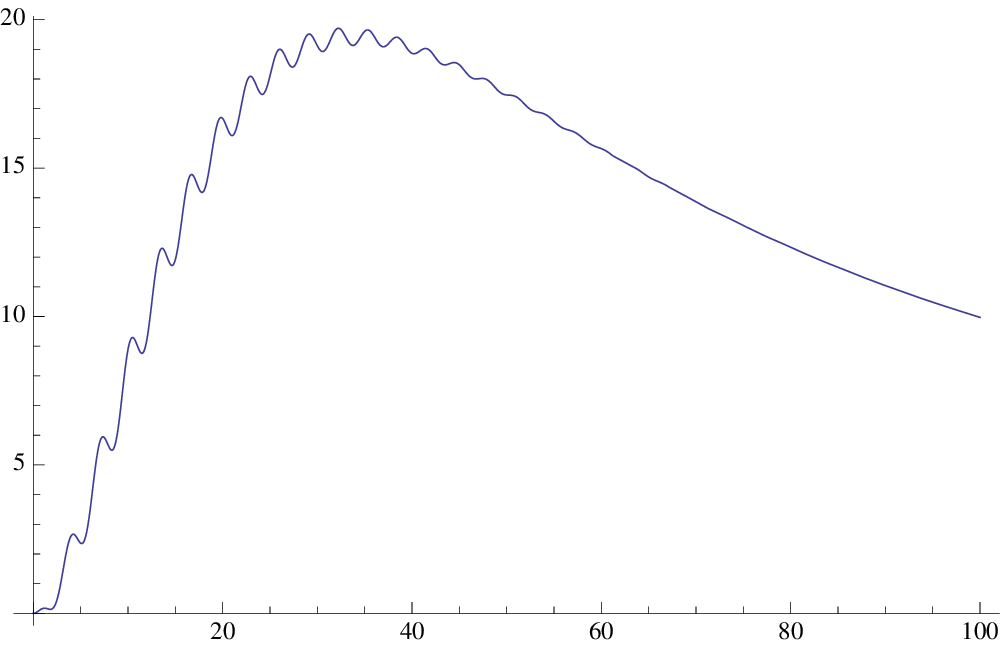}
\includegraphics[scale=0.62]{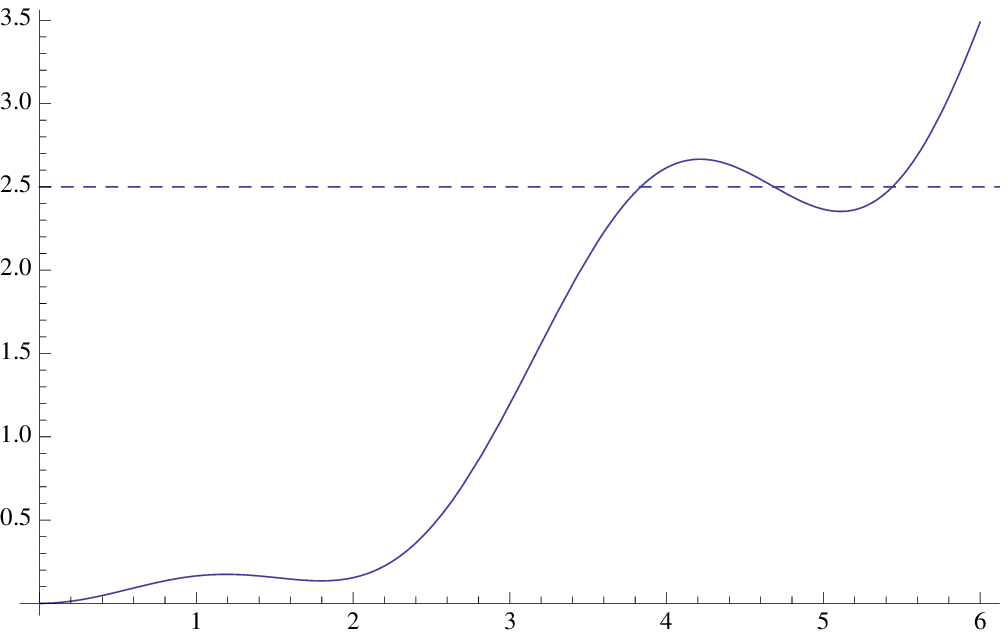}
\caption
    {\label{CE}
The left hand side of condition (\ref{con1}) for the distribution $\rho \sim k^2 \left( kx\cos^2(kx)\exp[-0.1kx]\right)$ as a function of $y=kx$. The top figure shows the condition globally in the whole range. At large enough $y$'s the left hand side of (\ref{con1}) decays to zero. The bottom plot, focuses close to the origin. The number of BHs is found by the number of intersection points of the plot with horizontal lines, which correspond to some fixed energy. Clearly, the horizontal (dashed) line with value $2.5$ intersects the curve three times at the vicinity of $y=4.5$ and (at least) one at infinity yielding four possible trapped horizons.}
\end{figure}

Relaxing the (iii) condition, which says that the expression $x(x\rho)'$ has a single root (other than maybe the root at $x=0$) and allowing more roots, generally yields more trapped surfaces. As an example, we consider the positive and integrable distribution $\rho \sim k^2 \left(kx\cos^2(kx)\exp[-0.1kx]\right)$ where $k$ is the transverse scale associated with the width of $\rho$. According to fig. \ref{VC}, it violates (iii) condition. Hence, more BHs are expected. Indeed, fig. \ref{VC} shows the violation of condition (iii) while fig.'s \ref{CE} indicate explicitly the possibility of more than two apparent horizons may be created.

An important observation is that at high enough energies, the number of the resulting trapped horizons is (at most) two and, this independently on whether assumption (iii) applies. This statement holds because (the left hand side of (\ref{con1})) vanishes only at $x=\infty$ and depending on the behavior of $\rho$ at the origin, occasionally at $x=0$. Lowering sufficiently the horizontal lines (see caption in fig. \ref{CE}) by increasing the energy, it will result in two intersection points (at most) and thus into two trapped horizons exhibiting some universal behavior independently on the (intermediate) details of the $\rho$'s.

\subsection{Thermodynamical implications}\label{3B}

We will discuss the small BHs of  \ref{3A3} with respect to their thermodynamical properties and thermodynamical stability. Most of the discussion of this part is based on assumptions and speculations and outlines several possible scenarios. However, a resolution of the open possibilities would be achieved when the actual Einstein's equations are solved and the problem is posed as an initial value problem in which case numerical relativity could say the final word. Our approach in this paper is simpler and inadequate in distinguishing the several ways of the evolution of the system, which will be mentioned below.

We begin from the Hawking formula for the entropy of a BH 
\begin{align}\label{S}
S_{BH}\geq S_{trap} =\frac{A_{trap}}{4G}= \frac{2}{4G}\int_0^{x_{\perp}^c}d^2 {\bf x_{\perp}}
=\frac{\pi}{G}(x^c_{\perp}(E))^2
\end{align} 
where A$_{trap}$ is the area of the trapped surface, the factor of 2 (second equality) comes from the two trapped surfaces $S_+$ and $S_-$ (see section (\ref{method})) while the dependence of the trapped horizon on the energy is explicitly given (third equality). In what follows we will assume $S_{BH}$ is proportional to $S_{trap}\equiv S$ and drop any subscripts on $S_{trap}$. Numerical works \cite{Wu:2011yd} (in AdS backgrounds) indicate that the difference is an energy independent proportionality factor.

We see two possible scenarios.\vspace{0.1in}

1. The co-existence scenario where the small surface is inside the big one (co-centric) and they are both created at the same time. Then, the following possibilities exist:
(a) Only the large BH mutters as the small one is hidden inside the large BH and its effects/presence will never be observed. (b) A related possibility is that the resulting black object will eventually evolve into a Reissner-Nordstrom (RN) BH. An RN BH has two co-eccentric horizons separating the BH in two regions. In this work, we find that the outer (inner) horizon increases (decreases) as the energy increases, similarly to the RN case. In addition there exists a critical value of the energy (see middle dashed line in fig. \ref{TSd}) where the two horizons coincide, resembling the extremal case in RN BH. The connection can be motivated further by defining the transverse scale of the width $k$ to be $r_Q^2 \equiv 1/k^2\equiv Q^2G$ where $Q$ a dimensionless parameter that can be dialed in independently on $E$. Hence, in our case as well, there also exist two length scales: $r_S=GE$ and $r_q^2=Q^2 G$ while the extremality occurs when $r_S/r_Q$ take a given minimum value. Decreasing the energy further, we find that a trapped surface is not formed. The corresponding picture in RN yields a naked singularity.

2. The separately formed scenario: the two solutions are possible saddle points of the (classical) theory. However since the large BH has a larger $S$, it has more probability to be formed. But in principle either the one or the other will be formed but not both (simultaneously; classically at least). We are mentioning the following possibilities: (a) The small BH is unstable and will decay immediately. (b) This (the small) solution naively\footnote{In the sense that at the instant of the collision, the system is far out of equilibrium and the discussion about temperature should be taken with caution.} yields negative temperatures because its size (and hence entropy) decreases as energy increases. hence, the small trapped horizon solution just doesn't make sense and it should be discarded.

\subsection{Universal behavior(s)}\label{UB}

In the high energy limit, the left hand side of (\ref{con1}) becomes $1/x_c$ because $x_c$ becomes large and as a result, $\int_0^{x_c} x \rho dx \rightarrow \sim 1$. Thus, $x_c \rightarrow \sim GE$ with the dependence on $k$ dropping out. This implies that in the high energy limit the following universal result\footnote{When two BHs exist, the universal behavior refers to the large BH.}, independent on the transverse distributions, applies 
\begin{align}\label{SH}
S\sim G(Ek)^2 \sim k^2s, \hspace{0.1in} Ek G\gg1.
\end{align} 

Another interesting point is to note that all the shocks in subsections \ref{3A1}-\ref{3A3} grow logarithmically at infinity. In fact, this is valid for any (positive, integrable) distribution $\rho$. To see this, one must note that the derivative of the shock is $\phi' \sim 1/x_{\perp} \int _0^{x_{\perp}} y\rho(y)dy$ and it becomes $\phi' \sim 1/x_{\perp}$ at large $x_{\perp}$. This implies that in 1+3 space-time dimensions, a (physical) distribution $\rho$, which creates a shock geometry such that it decays at infinity, does not exist. In $1+n$ space time dimensions the shock decays as $\phi \sim 1/x_{\perp}^{3-n}$ at large $x_{\perp}$. As we will see, there also exists the AdS version behavior of the shock-wave at large arguments and, which specifies the stress tensor in the gauge theory side.

\subsection{An application for LHC}\label{3C}

\begin{figure}
\includegraphics[scale=0.62]{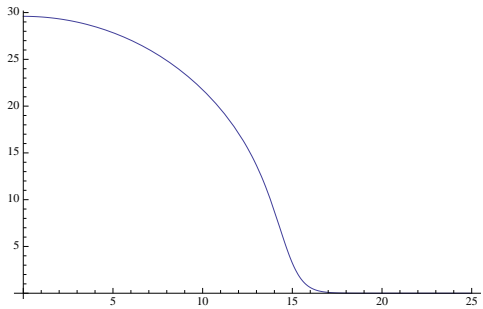}
\caption
    {\label{BWS}
     The (un-normalized) boosted Woods-Saxon distribution plotted numerically as a function of the transverse coordinates (in arbitrary units). Evidently this distribution belongs to the third class of distributions of this section and as a result, there expected two BHs if the energy is sufficiently large.}
\end{figure}

An interesting distribution to investigate is the Woods-Saxon (phenomenological) profile \cite{Woods:1954zz}, which seems to describe the nuclei distributions quite successfully. Boosting to the infinite frame by sending the boost factor to infinity but keeping the total energy $E$ fixed (at the order of the LHC scales), one obtains (see fig. \ref{BWS})

\begin{align}\label{WS}
T_{++}&=\frac{E}{c\sqrt{2}} \int_{-\infty}^{\infty} \frac{dx_3}{1+e^{\sqrt{(x_{\perp}/a)^2+(x_3/a)^2} -R/a}  }\delta(x^+)
\end{align} 
where $a \approx0.5$ fm, $R\approx 7.4$ fm is the Pb radius for the probes at LHC, $c=-\pi Li_3(exp({R/a)})$ fm$^2=13732$ fm$^2$ ($Li_3(x)$ is the polylogarithm function), which is chosen such that when (\ref{WS}) is integrated on the whole space to give the total energy $E$. The $\sqrt{2}$ is a consequence of our convention for the light-cone coordinates.

\begin{figure}
\includegraphics[scale=0.62]{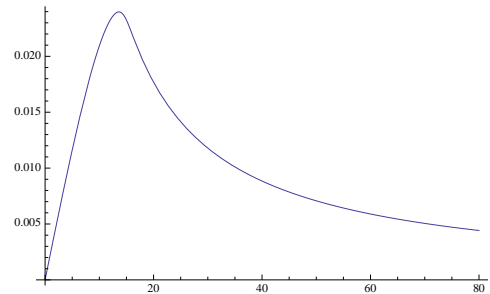}
\caption
    {\label{WS2}
     The (normalized) left hand side of (\ref{con1}) times $a=0.5$ fm plotted numerically as a function of the transverse coordinate (in units of fm). As expected from fig. (\ref{BWS}) and the discussion of subsection \ref{3A3}, two BHs may be formed given sufficiently large energy. The plot has a peak approximately at the point (13.5,0.024) and this point corresponds to the minimum energy where an apparent horizon could be created.}
\end{figure}

In fig. (\ref{WS2}) we plot the quantity on the left hand side of (\ref{con1}) by identifying $k$ with $1/a$ while both of the sides of the equation are dimensionless. As expected, it seems that two apparent horizons may exist, a small and a large one, given sufficiently large energy. 
On the other hand, the exact proportionality factor on the right hand side of (\ref{con1}) turns out to be $2\sqrt{2}/(16\pi)a \approx 0.03$ fm and as a result, this side of the equation is approximately equal to $0.03$ fm/GE. Converting everything in Planck mass units $(M_p^2=1/G)$ the right hand side of (\ref{con1}) becomes  
\begin{align}\label{PE}
1.8 \times 10^{18} M_p/E.
\end{align} 
Taking into account that, according to fig. \ref{WS2}, the trapped horizon with the least energy is formed when the left hand side of (\ref{con1}) is 0.024 and, equating this value with (\ref{PE}), it is deduced that the required energy is approximately $10^{20}$ orders of magnitude of the Planck mass, which is an extremely large energy. We conclude that, if this simplified model is not far from reality, no BHs will be seen at the LHC. However, there is a scenario discussed in the conclusions, which may avoid the obstacle in creating BHs of such a large scale set by the Newton's constant in 4 dimensions.

It is also interesting to note that any possible BHs found at the LHC (if the alternative scenario discussed in the conclusions is employed) could be of the RN type in the sense that they will consist of two co-eccentric horizons having a similar behavior with the energy as in the RN case. 

\section{$AdS_5$ background BHs production}\label{4}
Our investigation concerns head-on collisions of symmetrical distributions that (may) yield O(3) trapped geometries.  We classify the O(3) symmetric bulk distributions in AdS$_5$ according to whether trapped surfaces may or may not be formed. This symmetrical case is certainly not the most general one but it is convenient as it allows an analytical treatment of the problem and it suffices in showing our claims. The presentation of this section, which (initially) follows \cite{Gubser:2008pc}, will be brief as the ideas are similar with those of previous section. We begin from the analogue of equations (\ref{ds1}) and (\ref{Ein}) adapted to spherically symmetric shocks in AdS$_5$

\begin{align}\label{dsa}\hspace{-0.08in}
ds^2 &= b(z)^2\big(-2 dx^+  dx^- + d x_\perp^2 +dz^2\big) + b(z)\phi (q(x^1,x^2,z))\notag\\&
\times \delta(x^+)  d x^{+ \, 2}, \hspace{0.05in }b(z)=\frac{L}{z} , \hspace{0.05in }q=\frac{x_\perp^2+(z-z_0)^2}{4 z z_0}
\end{align}
where $L$ is the AdS$_5$ radius, $z$ is the fifth dimension with the boundary CFT residing at $z=0$, $q$ is the chordal distance in AdS$_3$, which makes the O(3) symmetry manifest while the role of $z_0$ is analyzed below. This ansatz, as before, satisfies trivially all the Einstein's equations except the $(++)$ component, which yields

\begin{align}\label{Eina}
&R_{++}+\frac{4}{L^2}g_{++}  = 8\pi G \left( J_{+ +} - \frac{2}{3} \, g_{++} J_{\mu}^{\mu}\right)\notag\\&
 => -1/2\delta(x^+)\left(\Box_{\perp}-3/L^2\right) \phi = 8\pi G (z/L)J_{++}  \notag\\&
 =8\pi EG{\tilde \rho} (q) \delta(x^+), \hspace{0.05in}J_{++}\equiv E (L /z) {\tilde \rho}(q) \delta(x^+)
\end{align}
where the $\Box_{\perp}$ operator is with respect to the $AdS_3$ metric 
\begin{align}\label{ads3}
ds^2=L^2\big[dq^2/(q(q+1))+4q(1+q)d\Omega^2\big],
\end{align}
G is the Newton's constant, which in this section concerns five dimensions, $E$ is proportional to the total bulk energy of the distribution, $z_0$ is the location of the center of the bulk distribution and, which according to the AdS/CFT dictionary (see \cite{Gubser:2008pc,Kiritsis:2011yn} for instance), is proportional to the width of the gauge theory stress-energy tensor (see also below). Both of the quantities $\phi$ and ${\tilde \rho}$ are functions of $q$ and they have dimensions of length to the first and the minus third power respectively. Dropping any positive dimensionless constants but keeping track of relative signs, previous equation yields the following differential equation for $\phi$
\begin{align}\label{dea}\hspace{-0.08in}
q(1+q)\phi''+3/2(1+2q)\phi'-3\phi& \sim -EG L^2 {\tilde \rho} \notag\\&
=-EG /L^2 z_0\rho
\end{align}
where $\rho \equiv L^4/z_o {\tilde \rho}$ as defined is dimensionless while $\phi$ has dimensions of length as it should. The choice of the overall coefficient for the bulk source comes about as follows: we need the field $\phi$ to expand at large $q$ as $1/q^3$ in order to ensure consistency with the AdS/CFT dictionary. Then, taking into account (\ref{dea}) and the fact that  $T_{++}\sim (L^2/G) \times 1/(zq)^3 |_{z=0}\delta(x^+)$ where $T_{++}$ is the gauge theory stress-tensor and assuming that we fix the integral of $T_{++}$ on the transverse plane to yield $E$\footnote{Allowing us to identify $E$ with the total collision energy in the gauge theory side. According to (\ref{Eina}), $E$ is (at the same time) proportional to the total bulk energy of the collision from the 5-dimensional point of view.}, we conclude that the overall dimension-full coefficient of $\phi$ is $EG/L^2 z_0$. Indeed, such a choice results $T_{++}\sim E \frac{zo^4}{(x_{\perp}^2+z_0^2)^3}\delta(x^+)$ which integrates to E.

The solution to equation (\ref{dea}) can take the integral form
\begin{align}\label{sol}\hspace{-0.08in}
\phi & \sim  E(G /L^2) z_0 \Bigg\{(1+2q)\left(C_1+\int ^q_{q_0}a_1(y)\int_{0}^y a_2(x)dxdy\right) \notag\\&
+C_2\left( \frac{-2(1+8q(1+q)}{\sqrt{q(1+q)}} +8(1+2q) \right) \Bigg\}, \notag\\& 
\hspace{0.25in}a_1(q)=\frac {1}{(1+2q)^2 (q(1+q))^{3/2}}\hspace{0.02in}, \notag\\&
 \hspace{0.25in}a_2(q)=(1+2q)\sqrt{q(1+q)} \rho(q)
\end{align}
where $C_1(q_0)$ and $C_2=\int a_2(y_0) dy_0\big|_{y_0=0}$ are constants chosen such that to remove the (1+2q) and $1/\sqrt{ q}$ terms of $\phi$ coming from the two lower limits of the integrations. This choice of $C_1$ yields (as mentioned above) $\phi \sim 1/q^3$ at large $q$ and ensures that the geometry, according to the AdS dictionary, corresponds to a boundary CFT theory.  $C_2$ on the other hand, should be chosen such that to cancel the $1/\sqrt{q}$ behavior of the integral at small $q$ because such a behavior induces an additional (to $\rho$) $\delta(q)$-source\footnote{Pf: 
In order to see this one should, using (\ref{ads3}), integrate $(\Box_{AdS_3}-3/L^2)f$ inside a small sphere of radius $q_0$ with $f\sim1/{\sqrt q}$ and finally take the limit $q_0 \rightarrow 0$: 
$\int_0 ^{q_0} \sqrt{g_3}\nabla^a\nabla_a \sqrt{q} dq d\Omega^2=\int_{S} \sqrt{g_2^{ind}}\left(\nabla_a \sqrt{q}\right) n^a d\Omega^2\big|_{q=q_0}=-1/2 \int_{S} q(1+q) q^{3/2} g_{qq}^{-1/2} d\Omega^2\big|_{q=q_0} \sim  \int_{S}  d \Omega^2 +O(\sqrt{q_0})\rightarrow \sim1$ where $g_2^{ind}$ is the induced metric of $g_3\equiv g_{AdS_3}$ on the surface $S: q=q_0$, $n^{\mu}$ a unit normal vector on  the same surface while after first equality, Gauss law has been employed.} on the right hand side of (\ref{dea}) (see \cite{Gubser:2008pc}). %
In particular, for distributions satisfying $\int _0^y a_2(y_0) dy_0\rightarrow 0$ as $y\rightarrow 0$, the coefficient $C_2=0$.

A few remarks are in order: (a) the first integral in (\ref{dea}) involving $a_2$ not only it should be finite as the upper limit tends to infinity but it should yield a positive constant. The reason is because, as we will see soon (see equation (\ref{con2})), it represents the bulk energy inside a ball of radius $y$ sitting in AdS$_3$ and this energy should be finite. This obviously occurs for a subset of choices of positive $\rho$'s; we call these physical $\rho$'s. (b) In particular, remark (a) implies that in the large upper limit expansion, this integral has a zeroth order term while the rest terms are sub-leading (involving negative powers). (c) The second integral involving $a_1$ when convoluted with negative powers and then expanded in the large upper $(q)$ limit, yields powers that drop faster than $1/q^3$. Hence, these terms do not contribute to the gauge theory stress-tensor. Convoluting with a zeroth power on the other hand, yields an $1/q^3$ fall-off and this term does contribute to the CFT stress-tensor. (d) Previous three remarks imply that physical sources yield a proper mapping of the geometrical description to the gauge theory description\footnote{The $1/q^3$ decay of $\phi$ at large $q$ can be justified rigorously as follows. The term in (\ref{sol}) which is proportional to $C_2$ decays as $1/q^3$ and so we ignore it from the further argument. Dividing the remaining terms with (1+2q), taking the derivative in both sides of the resulting equation and taking the large $q$ limit, taking into account that the integral involving $a_2$ is finite by assumption (see (\ref{con2}) and assumption (ii) in section \ref{4A}), yields $\left(\phi/(1+2q)\right)' \sim 1/q^5$ where $q\gg1$. Integrating $\left(\phi/(1+2q)\right)'$ (at large q) yields $\phi \sim 1/q^3$ up to $1+2q$ terms coming from the integration constants and which are cancelled by a suitable choice of $C_1$ in (\ref{sol}).}.


\subsection{General classification}\label{4A}


The analogue solution of (\ref{gc1}), to equation (\ref{ads}) with the condition $\psi(q_c)=0$ satisfied and where the curve $C: q=q^c$ defines the trapped horizon, is
\begin{align}\label{psa}
\psi(q)=\phi(q)-\frac{\phi(q_c)}{1+2q_c}(1+2q).
\end{align}

The derivative condition on the trapped surface in this case is $g^{qq}\psi \sim 1$ where $g^{qq}=q(1+q)/L$ is the (qq) component of the inverse metric of $AdS_3$. Then, employing (\ref{sol}) and (\ref{psa}) yields
\begin{align}\label{con2}
\frac{\int_0^{q_c}\sqrt{q(1+q)}(1+2q)\rho(q) }{(1+2q_c)q_c(1+q_c)} dq &\sim \frac{L}{z_0}\frac{L^2}{GE}\notag\\&
\sim \frac{L^3}{G}\frac{k}{E}, \hspace{0.08in} k\sim 1/z_0
\end{align}
where $k\sim 1/z_0$, as mentioned, is a transverse scale associated with the width of the gauge theory stress-tensor in the CFT. Equation (\ref{con2}) is the same as the condition derived in \cite{Gubser:2008pc} and has an analogous form and physical meaning as condition (\ref{con1}) with one crucial difference: the right hand sides of the two conditions differ by the fact that in (\ref{con1}) there exists the combination $E \times k$ while in (\ref{con2}) there exists the ratio\footnote{The dimensionless parameter $G_5/L^3$ is removed from the discussion as, in a bottom-up approach, it will be chosen such that it fits data \cite{Gubser:2008pc}.} $k/E$. This is a very important fact with potential consequences in heavy ion collisions and, which we will return  in section \ref{4C}.

The comments below condition (\ref{con1}) apply also here with minor modifications and so we will be brief in the subsequent presentation. We have therefore, all the machinery and the ideas in classifying the O(3) invariant distributions in AdS according to whether they produce one or more trapped surfaces during a collision. The distribution is again assumed (i) everywhere positive, (ii) integrable on the AdS$_3$ subspace and (iii) that the quantity $\sqrt{q(1+2q)}(1+2q)\rho(q)/(1+6q+6q^2)$ has at most one maximum.

 The consequences of these three assumptions are similar as in flat space (see discussion below (\ref{con1})). In particular, condition (iii) ensures that no more than two trapped surfaces may be created\footnote{Working analogously as in the flat space, one finds that a condition for having two trapped surfaces at most,  reduces in demanding that the function $g(q)=f/b$ where $f=\sqrt{q(1+2q)}(1+2q)\rho(q)$ and $b=\left(q(1+q)(1+2q)\right)'=(1+6q+6q^2)$ has (at most) a single maximum at some point $q=q_0$. Remark: The condition that the quantity $f=\sqrt{q(1+2q)}(1+2q) \rho$ has at most a single maximum is more physically motivated by analogy with the flat space and in the view of the integrand of (\ref{con2}). It is possible to make the two conditions applicable simultaneously by imposing extra conditions on $f$. For instance one could demand that $(\log(gb))'=(\log(f))'$ has only one root (at most) in the interval $(q_0,\infty)$ where $q_0$ is the (only possible) maximum of $g$. In fact, the two conditions happened to coincide for all the cases that we have checked.}.

\subsubsection{Always a single black hole for any energy }\label{4A1}

This case corresponds to distributions that at small $q$ behave as $\rho \sim1/q^n$+(sub-leading) where $3/2>n>1/2$. The upper bound comes by demanding finite total energy while the lower bound comes by demanding the left hand side of (\ref{con1}) to diverge as $qc\rightarrow0$.

As an example of such a distribution one may consider $\rho \sim 1/[q(1+2q)(1+q)^2]$, which as can be checked, creates the shock $\phi \sim -EGz_0/L^2(1 + 2 q) \left(1/(1 + q) + 4/(1 + 2 q) + \log[q(1+2q)^4/(16(1+q)^5)]\right) $. At large $q$ this shock-geometry behaves as $1/q^3$, which implies the stress-tensor in gauge theory behaves as $E/(x_{\perp}^2+z_0^2)^3$ and is also positive as should. It is pointed that the relative signs work out\footnote{Such that both, the (dual) gauge theory stress-tensor and the bulk tensor are positive.} and that  $z_0$ is the transverse width of the gauge theory stress-tensor. The left hand side of (\ref{con2}) behaves as $1/\left({\sqrt q}(1 + q)^{3/2} (1 + 2 q)\right)$ and the corresponding plot topologically looks like fig. \ref{I1} showing that a single trapped surface is always formed for any energy. Evidently, at large $E$ one obtains $q_c^3 \sim E/k$.

\subsubsection{Marginal case: one black hole for sufficiently large energy }\label{4A2}

This case corresponds to distributions that at small $q$ behave as $\rho \sim1/\sqrt{q}$+(sub-leading). As an example of such a distribution one may consider $\rho \sim 1/[(1 + 2 q)^2 \sqrt{q} (1 + q)^{3/2}]$, which as can be checked, creates the shock $\phi \sim EGz_0/L^2$ $\Big(2 q -4\sqrt{ q (1 + q)} (1 + 2 q)\left( \tan^{-1}\left( \sqrt{\frac{q}{1 + q}}\right)+ \log(2)-\pi /4 \right) $
$ - (1 + 8 q (1 + q)) \log\left((1 + q)/(1 + 2 q)\right)\Big)/ \sqrt{q (1 + q)} $. At large $q$ the shock decays as $1/q^3$, which is in accordance with the AdS dictionary while, the left hand side of (\ref{con2}) becomes $\sim \log[(1 + 2 q)/(1 + q)]/(q(1+q)(1+2q))$, which vanishes as $q \rightarrow \infty$ and it is finite when $q=0$. Topologically, this plot is like that one of fig. \ref{I2} showing that only a single trapped surface may be formed for sufficiently large energy. In addition, at large $E$ one again obtains $q_c^3 \sim E/k$ as in \ref{4A1}.

\subsubsection{Two black holes for sufficiently large energy}\label{4A3}

This case corresponds to distributions that at small $q$ behave as $\rho \sim 1/q^n$+(sub-leading) where $n<1/2$. As an example of such a distribution one may consider the $n=-1/2$ and in particular the distribution $\rho \sim  \sqrt{q}/\left((1+q)^{5/2}(1+2q)^2\right)$, which as can be checked, creates the shock $\phi \sim -EGz_0/L^2 \Big(3 + 2 q (30 + q (61 + 32
 q)) - 36 (1 + q) \sqrt{q (1 + q)} (1 + 2 q) \tan^{-1}\left(\sqrt{q/(1 + q)}\right)+ 
  6 (1 + q) (1 + 8 q (1 + q)) \tanh^{-1}\left(q/(2 + 3 q)\right) \Big)/\left( 3\sqrt{q} (1 + q)^{3/2}\right)+EGz_0/L^2 \Big((1 + 8 q + 8 q^2) / \sqrt{q (1 + q)}\Big) +EGz_0/L^2\Big((4/3 (5 + \log(8)) - 3 \pi)$ $(1 + 2 q)\Big)$. The second and first terms from the end subtract the delta function source at small $q$ and ensure a $1/q^3$ (in order to have a proper CFT mapping) fall-off at large $q$ respectively. Remarkably, this shock-geometry is everywhere finite, has finite first order derivatives but irregular second and higher order derivatives at $q=0$.

 The left hand side of (\ref{con2}) becomes $\sim \Big(q + (1 + q) \log \left[(1 + q)/(1 + 2 q)\right]\Big)/\left(q (1 + q)^2 (1 + 2 q)\right)$, which vanishes as $q\rightarrow 0$ and as $q \rightarrow \infty$. Topologically, this plot is like that one of fig. \ref{TSd} showing the possibility of creation not of just one but of two trapped surfaces, a small and a large one, for sufficiently large energy. Finally, the behavior of $q_c$ with $E$ in the high energy limit is the same as in previous cases, namely as $q_c^3 \sim E/k$. 
 
One should bare in mind the analytical work of \cite{Dias:2011ss} and the numerical results of \cite{Bizon:2011gg,Jalmuzna:2011qw} concerning AdS$_5$ background. In these works, a BH eventually, after multiple re-scatterings with the boundary of the AdS, will be formed. However, these works have a different set-up from the current one. They use a global covering of the AdS$_5$ space, the corresponding bulk fields are self interacting in the sense that they obey their own equation of motion and most importantly, these are exact conformally invariant set-ups. In the contrary, in our case the presence of the energy $E$ and the transverse width $k$ of the distribution breaks conformality (softly). Hence the fact that a BH is not always formed is a possible scenario as supported by our results and also from other works \cite{Dias:2012tq}.
 \subsection{Thermodynamical implications}\label{4B} 

 For AdS backgrounds we know that the AdS-Schwarzschild large BH is thermodynamically stable and the small one is unstable. Thus, it seems that all the trapped horizons that we find and, which eventually will form BHs, will be
 probably thermodynamically stable except maybe from the small horizons of subsection \ref{4A3} (in a non-coexisting scenario see section \ref{3A3}). In the coexisting scenario, where both, the large trapped surface surrounds the small one, it is possible that the system will evolve into an AdS-RN BH where, as in the flat space, the third required scale (in addition to the AdS$_{5}$ radius and the Schwarzschild scale due to $E$) is probably the transverse width of the distribution.

\subsection{Implications for heavy ion collisions and QGP}\label{4C}

At RHIC and now at LHC the QGP production is expected that is formed during heavy ion collisions. AdS/CFT has been proven a powerful tool in modeling these kind of collisions \cite{Albacete:2008ze,Lin:2010cb,Taliotis:2009ne,DuenasVidal:2012sa,Albacete:2008vs,Albacete:2009ji,Taliotis:2010pi,Wu:2011yd,Grumiller:2008va,Bantilan:2012vu,Heller:2012je,Wu:2012rib,Arefeva:2012jp,Erdmenger:2012xu,Baier:2012ax,Galante:2012pv,Caceres:2012em,Caceres:2012ii,Dusling:2008tg,Kovtun:2004de,Policastro:2001yc,Buchel:2012gw,Buchel:2008vz,CasalderreySolana:2011us,Balasubramanian:2010ce,Balasubramanian:2011ur,Balasubramanian:2011at} and predicting the QGP formation, which in the AdS/CFT context, implies a BH creation. However, the existing literature has missed one important feature of QGP observed experimentally and, which is according to our intuition: QGP, even for very central collisions, is formed for sufficiently large energies, much larger than $\Lambda_{QCD}$. However, the existing works have found that the criterion in forming a BH shares qualitatively the features of the plot in fig. \ref{I1}: a BH is formed for any small energy (as long as the collision is central).

Identifying $k$ with $\Lambda_{QCD}$ in (\ref{con2}) is now tempting. According to (the analogues of) fig.'s \ref{I1} and \ref{I2} (adapted for the AdS$_{5}$ background case), it is deduced that at high energies such that $E \gg k=\Lambda_{QCD}$, there are more chances to get a BH and hence to produce QGP. Opposite conclusions apply in the $E\ll k$ limit. This is exactly what it is expected/observed in realistic heavy ion collisions and it is pleasing that it can be modeled geometrically.

In this work, we have shown that one can built geometries that both, map properly onto a CFT and, which when applied in heavy ion collisions, they capture the prescribed feature of QGP formation. We assumed an O(3) symmetry for simplicity, which allowed us to derive our conclusions using purely analytical methods. This symmetry, which is not necessarily realistic (in real heavy ion collisions; see \cite{Gubser:2011qva} however) together with (approximate) conformal invariance, force the gauge theory stress-tensor of the colliding matter to behave as a power law and in particular as $\sim Ez_0^4/(x_{\perp}^2+z_0^2)^3\delta(x^+)$. In addition, as we have seen from all the examples we have studied, in the high energy limit $q_c \sim (E/k)^{1/3}$. This implies that total multiplicities N grow as N$\sim S\sim (s/k^2)^{1/3}$ where $s$ is the c.m. energy of the heavy ions beam\footnote{The fact that $N\sim S$ is based on phenomenological assumptions and it is not straightforward. The fact that $S\sim s^{1/3}$ will be explained shortly. We consult the reader to see \cite{Kovchegov:2009du,Lin:2009pn,Gubser:2008pc,Arefeva:2012ar,Kiritsis:2011yn}.}. This is a general feature of the O(3) symmetry and yields a universal behavior\footnote{Departing from the O(3) symmetry, different (than the power law) gauge theory stress-tensors become possible and one may find a model that is closer to more realistic nuclear profiles. To this end, multiplicities are generally going to be different.} for $S(E)$ similar to the flat space of section \ref{UB}. Such a universal high energy limit behavior predicts a slightly larger growth of multiplicities compared to the data ($N\sim s^{1/4}$)\footnote{See \cite{Kiritsis:2011yn,Stoffers:2012mn} for a more involved treatment where a fitting with RHIC and LHC data as well as future predictions are achieved.}. In order to see this universal result, one has to take the high energy limit and hence the large $q_c$ limit of (\ref{con2}) taking into account that the distribution $\rho$ is integrable (by physical assumption). Hence the numerator of the left hand side of this equation approaches a constant and the equation yields $q_c^3\sim E/k$. The final step is to compute $S$ from the adaption of (\ref{S}) for AdS$_5$ with the subspace measure (that of AdS$_3$). At large $q_c$ it yields $S\sim q_c^{2/3}$ and hence $S\sim N\sim (s/k^2)^{1/3}$.


\subsubsection*{Attempting to fit the RHIC and the LHC data by incorpo- rating weak coupling physics: a phenomenological approach}

\begin{figure}
\includegraphics[scale=0.82]{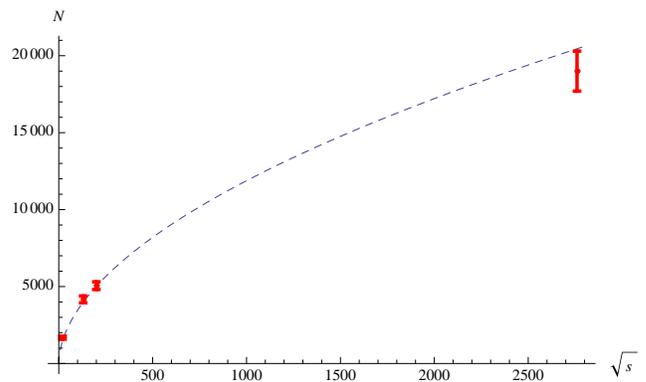}
\caption
    {\label{N}      
    The PHOBOS \cite{Back:2002wb} and the preliminary ALICE \cite{Collaboration:2011rta} (right point) data including the error bars. The vertical axis denotes the total multiplicity while the horizontal denotes the  nuclon-nucleon c.m. energy meassured in GeV. The far right (ALICE) point has been adjusted from \cite{Collaboration:2011rta} by a factor of $2A_{Pb}/375$, where $A_{Pb}$ is the atomic number of Pb, in order to compensate the non-exact centrality aspect of the process. The dashed line is our theoretical prediction, (formula (\ref{Ns})) $N=295 (\sqrt{s}/GeV)^{0.54}$ where the number $295$ is the only (energy independent) fitting factor. 
}
\end{figure}

It is possible that the scale $k$ could depend on the energy. Instead of identifying the transverse scale of the gauge theory stress-tensor $k$ with $\Lambda_{QCD}$ one could identify it with $Q_s(E)$ that is with the saturation scale \cite{JalilianMarian:2005jf}. Such an identification is physically motivated as $Q_s$ is the typical transverse momentum of the gluon field (which dominates over the fermions) in a highly boosted nucleus. Furthermore, taking $k=Q_s$, incorporates, in a sense, the weak coupling physics that also participate in heavy ions collisions. Taking into account that perturbative calculations \cite{Triantafyllopoulos:2002nz,Gelis:2010nm,Levin:2011hr,Gotsman:2002yy} yield $Q_s\sim \Lambda_{QCD} (\sqrt{s})^{\lambda}$, $\lambda \in [0.1,0.19]$ with $\lambda=0.19$ \cite{Gotsman:2002yy} referring to nucleus-nucleus collisions and $s$ is measured in GeV, it is deduced that\footnote{Equation (\ref{Ns}) may have significant corrections due to both, higher order corrections and also running coupling effects of $\lambda$ \cite{Albacete:2007yr}.}
\vspace{-0.1in}
\begin{align}\label{Ns}
N \sim \left(\frac{s}{Q_s^2(s)}\right)^{1/3} \approx \hspace{0.1in} \sim (s /\Lambda^2_{QCD})^{0.27}.
\end{align}
Such a dependence of $N$ on $s$, which is very close to the commonly accepted power $N\sim s^{1/4}$ (see \cite{Kiritsis:2011yn} and references therein), may fit the Phobos \cite{Back:2002wb} and the first ALICE \cite{Collaboration:2011rta} data satisfactory within the error bars, up to an overall energy independent proportionality factor (see fig. \ref{N} and \cite{Kiritsis:2011yn} for a different but related approach). This result is universal, independent on the details of the transverse distributions, as long as the high energy limit is assumed. This implies that $E=1/2\sqrt{s}\gg k=Q_s$ should apply, which means $({\sqrt s}/\Lambda_{QCD})^{1-\lambda}\gg1$ should hold and, which is a consistent inequality for the energies at RHIC and at LHC.

\subsubsection*{Mapping the gravity evolution to the evolution of the produced medium in heavy ions}

A possible evolution of the system in gravity and its implications in heavy ions motivated from the gravity analysis of section \ref{4A3} is the following.  After the shocks collide, two co-eccentric 3-dimensional black disks, the trapped surface, is formed. This surface has (almost) a zero depth in the longitudinal direction. The projection of this configuration in $R^{1,3}$ is 2-dimensional disk, that is a 2-dimensional plasma. This applies immediately after the collision. Then, the trapped horizon will isotropize due to the huge longitudinal pressure gradient in the center of the black object and it will evolve into a (RN) BH; exactly as it is expected in heavy ions \cite{Kovchegov:2005az,Kovchegov:2005ss,Heller:2012km} where the resulting thermal medium becomes isotropic. Such an RN-like configuration could capture the presence of a chemical potential \cite{Arefeva:2012ar} in QGP. Since the inner horizon of an RN BH has different thermodynamical properties than the intermediate shell extending to the outer horizon \cite{Balasubramanian:2004zu}, this system could model a hotter central region and a thermodynamically unstable periphery \cite{Gubser:2000ec,Konoplya:2008rq,Chamblin:1999hg,Louko:1996dw} during the medium's last stages of evolution. Finally, the BH will undergo through a Hawking evaporation, which implies a hadronic phase transition in the heavy ions picture. This difference in the thermodynamical behavior of the inner-outer region is qualitatively in accordance with the hydrodynamical simulations \cite{Kolb:2003dz,Huovinen:2001cy,Schnedermann:1993ws} of the QGP evolution.

\section{Summary, Conclussions and future directions} \label{conc}


In this work we have studied the problem of trapped surface(s) formation resulted by colliding extended ``spherically" symmetric, physical sources on the transverse plane and boosted to the speed of light. The total energy of the collision has been assumed finite and the impact parameter zero. The investigation has taken place in 4 dimensional flat and AdS$_5$ backgrounds. 

 \vspace{0.1in}

Our conclusions are summarized as follows.

  \vspace{0.1in}

I. For both backgrounds the spherically symmetric distributions have been classified according to: (i) always a single apparent horizon for any $E$ (ii) a single apparent horizon for sufficiently large $E$ (iii) two apparent horizons for sufficiently large $E$. Last category suggests that small and large BHs may be formed even in flat backgrounds, which, at the best of our knowledge, is a new result.

 \vspace{0.15in}

II. We found that the creation of small BHs in both types of backgrounds is possible. In section \ref{3B} we have presented the possible scenarios for the flat backgrounds with similar possibilities for the AdS backgrounds. We find particularly attractive the suggestion that the resulting configuration will finally equilibrate into an RN BH. The role of the scale multiplying the charge in the RN BH should be played by the transverse scale of the distribution. In particular, considering AdS$_5$ backgrounds, the RN scenario would explain the appearance of a chemical potential during heavy ion collisions. Such a configuration could model a hotter core (using the inner horizon) having different thermodynamical properties than the outer region (the shell between the inner and the outer horizon) as it happens in realistic collisions, according to hydrodynamical simulations \cite{Kolb:2003dz,Huovinen:2001cy,Schnedermann:1993ws}. A detailed scenario based on the RN BH explanation, with emphasis in the implications of the QGP evolution, is proposed at the end of section 
\ref{4C}. In any case, all scenarios worth further investigation while possibly only numerical relativity may answer these sort of questions reliably.

 \vspace{0.15in}

III. There is a universal behavior independent on the colliding bulk distributions, which have a transverse scale $k$ that generally depends on $E$. In the high energy limit, one obtains $S\sim (kE)^2$ and $S\sim (E/k)^{2/3}$ for flat and AdS$_5$ backgrounds respectively for any collided distribution obeying standard assumptions found in sections \ref{3}, \ref{4}. This applies even when assumption (iii) (see sections \ref{3} and \ref{4}) is relaxed. In the introduction, it was claimed that this result was independent on the presence of any extra dimensions and on the impact parameter $b$ as long as the energy is sufficiently high. For the extra dimensions, this was shown in \cite{Taliotis:2012qg} in higher dimensions. The idea is clear: at very large energies, the resulting black object occupies the whole compact directions and becomes independent on them (see for instance (\ref{cr}) for large $x_{\perp}$). An analogous idea motivated by the exact result of \cite{Eardley:2002re}, applies for having a non-zero $b$ which results in a non-zero angular momentum and yields an elliptical trapped horizon. However, as $Eb \rightarrow \infty$, the surface becomes (more) circular. 

 \vspace{0.1in}

IV. A phenomenological model for describing possible (4-dimensional) BH creation at the LHC has been constructed using the Woods-saxon profile as the colliding matter. This profile results to two trapped horizons given sufficiently large energy (see fig. \ref{WS2}). Substituting the numbers, it is found that the minimum energy required is of the order of $10^{20}$ of $M_p$, pointing to the direction that no BHs will ever be seen at the LHC. This result is consistent with the fact that the SM interactions inside the world as we know it today, dominate over the gravitational ones for energies way below $M_p$. However, as it is discussed below, there might be a way around. 

 \vspace{0.1in}

V. Classes (ii) and (iii) of the first conclusion (see above) show the possibility in building geometries that both, mimic heavy ion collisions and,  which predict QGP formation only for sufficiently large energies compared to the transverse scale involved and, which is tempted to be interpreted as the $\Lambda_{QCD}$ scale. This fact applies (even) for perfectly central collisions. In addition, the O(3) geometries seem to predict that the total multiplicities at high energies behave as $(s/k^2)^{1/3}$ independently on the details of the profiles of the  bulk distributions.

 \vspace{0.1in}

VI. In the AdS$_5$ space, the transverse scale of the bulk distribution coincides with the transverse width of the gauge theory stress tensor before the collision and, which in turn, it can be thought as a nucleus moving with the speed of light. In the color glass condensate language (CGC) \cite{JalilianMarian:2005jf,McLerran:1993ni} this width is the saturation scale $Q_s$. It is of the order of GeV and it depends on the energy in a fashion which is computed perturbatively \cite{Mueller:2002zm}. Identifying $k$ with $Q_s$ is not only a plausible assumption but it also incorporates in a sense, the weak coupling physics in the large coupling treatment that AdS/CFT relies on. Such an identification yields the result that total multiplicities grow as $N\sim (s/{\Lambda_{QCD}^2})^{0.27}$ which seems to describe satisfactory the PHOBOS \cite{Back:2002wb} and the ALICE \cite{Collaboration:2011rta} data at RHIC and at LHC respectively. The corresponding plot is depicted in figure \ref{N}.

 \vspace{0.1in}

VII. A number of investigations have been postponed for future work. The non-zero impact parameter case corresponds to more realistic situations, especially in heavy ion collisions (i.e. elliptic flow \cite{Kolb:2003dz}) and should be considered. However, we expect that it will not change the results in the high energy limit (see conclusion IV and the results of \cite{Eardley:2002re,Gubser:2009sx} in the limit $GE/b\gg1$).

\subsubsection*{Introducing extra dimensions and BHs at the TeV scale}

\vspace{0.2in}

VIII. A more interesting new direction is the generalization to more dimensions.  Although a straightforward  task, such a generalization will turn out useful in taking another step further: the inclusion of extra, flat dimensions for flat backgrounds, which is a valid scenario for the BHs production in the accelerators as it could lower the Planck mass and make the production of BHs feasible even at TeV scales. Related to this, we make the following hypothesis followed by a conjecture based purely on dimensional analysis. We require the creation of BHs at the LHC for collision energies of a few TeV's in the presence of extra dimensions having magnitude $R_i$, $i=1,2,...n$. We also assume that the colliding matter from a 4 dimensional point of view, belongs to the class of distributions of subsection \ref{3A3}, which are the more realistic ones and, which allow a BH creation only if the energy is above a threshold that typically is of the order of the Planck mass ($M_p$) or beyond. We then conjecture that creating BHs at lower scales than $M_p$ is possible if the width $r_i\equiv 1/k_{R_i}$ of the matter in the extra directions is much smaller than the size $R_i$ of the compact dimension(s). In other words BHs could be created at the LHC when the matter is sufficiently localized in the extra dimensions such that $k_{R_i} R_i\gg1$ is satisfied.
Argument: since the Newton's constant when $n$ compact dimensions are present, is $G_{4+n}=G_4 R_1...R_n$, then the dimensionless coupling controlling the creation of the BHs should be
\begin{align}\label{conj1}
w&\equiv  \frac{1}{G_{4+n}}\frac{1}{ Ek k_{R_1}...k_{R_n}}=\frac{M_p^2}{Ek}\frac{1}{ (k_{R_1}R_1)...(k_{R_n} R_n)}\notag\\&
=\frac{M^2_{eff}}{Ek},\hspace{0.02in}M^2_{eff} \equiv \frac{M_p^2}{ (k_{R_1}R_1)...(k_{R_n} R_n)},\hspace{0.02in}G_4=M_p^{-2}
\end{align}
where $E$ and $k$ are the total energy and the transverse width from the 4 dimensional point of view while $M_{eff}$ is an effective Planck mass. Hence, for $w\sim 1$ and taking into account that $M_p^2/(Ek)\gg1$, then the $R/r_{k_i}=k_{R_i}R_i\gg 1$ condition follows. This condition should had been expected. The reason being when this strong inequality is satisfied, the matter is spread along the 3-brane but it is localized (almost like a delta function) in the extra directions. But a delta function distribution belongs to the first class (see subsection \ref{3A1}) of distributions, which yield a BH for any small collision energy. Therefore, it seems that the localization of the matter in the extra dimensions, compensates the (realistic) spread in the two (transverse) extended dimensions allowing the possibility in creating BHs at lower scales.

Taking $E\sim 10$ TeV and $k \sim 1$ GeV equation (\ref{conj1}) takes the suggestive form
\begin{align}\label{conj2}
w=\frac{M^2_{eff}}{M^2_{EW}},\hspace{0.02in}M^2_{EW} \equiv Ek\sim(100 \hspace{0.02in}{\mbox GeV})^2
\end{align}
where $M_{EW}$ is of the order of the electroweak scale. In particular, for $n=2$, $R_1=R_2=R\sim 0.1$ mm as in \cite{ArkaniHamed:1998rs,Antoniadis:1998ig} and demanding $w\sim1$ it is deduced that $k_{R_1}=k_{R_2}=k_R\sim10^{18}/\mbox{mm}=10^{30}$/fm$\sim 10^{30}$ GeV and so $r_{k_R}=1/k_{R}\sim 10^{-30}$ fm.

It is remarked that such a (small) spread \cite{Frere:2003hn,ArkaniHamed:1999dc,ArkaniHamed:1999za,Cacciapaglia:2007fw} in the compact directions not only is allowed, as it does not violate anything fundamental, but is also essential from a quantum uncertainty point of view and thinking about the problem semi-classically. It is experimentally allowed for SM interactions to penetrate the extra dimensions as long as the corresponding penetration length\footnote{From a 4 dimensional point view such a characteristic length can be taken as the proton's radius. In the extra dimensions scenario, the length is the width of the distribution.} is much smaller than the resolution of the accelerators .The energy required to resolve such small widths are of the order of the Planck mass or even (as in the example above) way beyond. Hence such a small penetration depth is practically unobservable.

As a candidate distribution, which could model the problem of not creating BHs in the absence of extra dimensions at a given energy but with the possibility of BHs creation in the presence of extra dimensions at that same energy, one could use the distribution
\begin{align}\label{cr}
\rho = \frac{k^2k_R^2 \exp \left[-k^2x_{\perp}^2-k_R^2R^2\left(\sin^2(\theta_1/2)+\sin^2(\theta_2/2)\right)\right]}{4\pi^3  \left (k_R^2R^2 \right)e^{-k_R^2R^2}I^2_0(\frac{1}{2}k_R^2R^2)}.
\end{align}
This distribution, which is normalized to unity and where $I_0$ is a modified Bessel function, has the following features: (i) It is periodic with respect to the angles $\theta_1$ and $\theta_2$ that parametrize the extra dimensions and it peaks in the vicinity of $\theta_1=\theta_2=2n\pi$, $n\in Z$ where SM resides\footnote{Modulo a small but unobservable penetration inside the compact directions.}. (ii) Ignoring the extra dimensions, it belongs to the class of distributions of \ref{3A3} where two BHs will be created for sufficiently large energy. This can be seen if one integrates out the compact coordinates. Then, the resulting curve is a gaussian and topologically it is what one expects for realistic distributions in the 4 dimensional point of view (it looks like the one of fig.\ref{BWS}). This is what  happens at large energies. The trapped surface becomes very large and covers the whole compact directions \cite{Taliotis:2012qg} (it becomes a black string) and the distribution effectively becomes 2 dimensional depending only on the transverse coordinate $x_{\perp}$. (iii) When the collision energy is sufficiently small, the trapped surface is too small to resolve the finiteness of the extra dimensions (see \cite{Taliotis:2012qg} for a detailed analysis on this). Thus, when the the distribution is expanded (at small angles and small $x_{\perp}$), yields a rotationally invariant expression if in addition, $k_R=4k$ is assumed. Then, the problem in this energy regime can be solved exactly and this would provide cross-checks\footnote{Along with the high energy limit-see (ii) above.} of a more thorough (numerical) analysis. (iv) Plotting the distribution for fixed $x_{\perp}$, it looks very localized; almost like a delta function. In fact, in the limit $k_R R\rightarrow \infty$ it becomes a delta function while the denominator of (\ref{cr}) becomes $k_RR$ independent.

This distribution, from a 4 dimensional point of view (so ignoring the angles), creates BHs when $Mp^2/(Ek)\sim1$ is fulfilled. The question is whether the introduction of the extra dimensions could yield BHs even when $Mp^2/(Ek)\gg1$ applies. The problem of solving the underlying trapped surface equations could be attacked numerically and answer quantitatively these questions. However, the intuition we obtained suggests that it is possible to have BHs in the $Mp^2/(Ek)\gg1$ regime as the presence of the extra dimensions could lower $M_p$ down to $M_{eff} \sim M_{EW}$ yielding a BH at accessible collision energies.  Hence, under the assumptions that extra dimensions do exist and that the matter is dense enough in those directions, implies that LHC is maybe not that safe after all.


\acknowledgments

The author is supported in part by the Belgian Federal Science Policy Office through the Interuniversity Attraction Pole P7/37, and in part by the ``FWO-Vlaanderen" through the project G.0114.10N, which he acknowledges. Acknowledge is also granted to the colleagues: V. Balasubramanian, J. de Boer, B. Craps, J.M. Frere, F. Galli, S. Giddings, D. Grumiller, U. Gursoy, G. Horowitz, A. Leibman, M. Lublinsky, S. Mathur, E. Kiritsis, Y. Kovchegov, P. Romatschke, A. Rostworowski, D. Triantafylopoulos and G. Veneziano for valuable discussions. In addition, we thank members of the Niels Bohr Institute and the organizers of  ``Black Holes and Applied Holography" and especially J. Hartong and N. Obers for their warm hospitality during the final stages of this work.





\bibliography{references}
\bibliographystyle{unsrt}
\end{document}

\end{document}